\documentclass[submission,copyright,creativecommons]{eptcs}
\providecommand{\event}{ASQAP 2025} 

\usepackage{iftex}
\usepackage{xcolor}
\usepackage{graphicx}
\usepackage{subcaption}
\ifpdf
  \usepackage{underscore}         
  \usepackage[T1]{fontenc}        
\else
  \usepackage{breakurl}           
\fi

\title{Towards Federated Digital Twin Platforms}
\author{Mirgita Frasheri \thanks{All authors acknowledge the Poul Due Jensen Foundation for 
supporting Digital Twin Research at Aarhus University.}
\institute{Department of Electrical and Computer Engineering \\ DIGIT, Aarhus University}
\email{mirgita.frasheri@ece.au.dk}
\and
Prasad Talasila
\institute{Department of Electrical and Computer Engineering \\ DIGIT, Aarhus University}
\email{prasad.talasila@ece.au.dk}
\and
Vanessa Scherma
\institute{Politecnico Di Torino}
\email{vanessascherma.job@gmail.com}
}
\def\titlerunning{Towards Federated Digital Twin Platforms}
\def\authorrunning{M. Frasheri et al.}
\begin{document}
\maketitle

\begin{abstract}
Digital Twin (DT) technology has become rather popular in recent years, promising to optimize production processes, manage the operation of cyber-physical systems, with an impact spanning across multiple application domains (e.g., manufacturing, robotics, space etc.). 
DTs can include different kinds of assets, e.g., models, data, which could potentially be reused across DT projects by multiple users, directly affecting development costs, as well as enabling collaboration and further development of these assets.
To provide user support for these purposes, dedicated DT frameworks and platforms are required, that take into account user needs, providing the infrastructure and building blocks for DT development and management. 
In this demo paper, we show how the DT as a Service (DTaaS) platform has been extended to enable a federated approach to DT development and management, that allows multiple users across multiple instances of DTaaS to discover, reuse, reconfigure, and modify existing DT assets.
\end{abstract}

\section{Introduction}
Interest in Digital Twins (DTs) has soared in the past years, with the techonology promising to optimize 
production processes, as well as delivering run-time services such as monitoring, fault detection
and diagnosis, as well as reconfiguration of processes or systems~\cite{fitzgeraldEngineeringDigitalTwins2024}. 
A DT is usually defined as a virtual replica of a physical or cyber-physical system, referred to as the
physical twin (PT), with continuous bi-directional communication between the two~\cite{Kritzinger2018DTs}.
As such, the operation of the DT is impacted by the real data coming from the PT, and in turn the operation
and behaviour of the PT is influenced by the analyses and services provided by the DT.
Note however, that a DT is not to substitute the functionality of the PT, but rather enhance it, and in this
case we can refer to the PT as a DT-enabled system~\cite{fitzgeraldEngineeringDigitalTwins2024}.
DTs are useful across application domains, from manufacturing, to robotics and space, in monitoring
the degradation of materials, or possible mission failures, and executing viable mitigation strategies
to avoid failures or extended down-times.

In recent years, the focus of the DT research community has started to shift from the tools and methodologies
for building individual DTs in rather ad-hoc ways, to platforms and frameworks that support the creation,
execution, and more generally the management of possibly different DTs, as well as providing configurable
services that enable DT-PT communication~\cite{gil2024survey}.
In addition, platforms such as the Digital Twin as a Service (DTaaS)~\cite{talasila2024composable} allow users to share DTs, as well
as other finer grained assets, e.g., models, tools, and data, thus making it possible for users to resuse
these across different DT applications and users themselves. 
Nevertheless, the collaboration of users has remained at a higher level of abstraction, e.g., in
terms of exchanging plug-and-play models, or tools. The flexible reconfiguration and execution of DTs is not addressed to a satisfactory extent.
To address this issue, DTaaS has been extended to integrate the Gitlab DevOps (combination of development and operation activities) framework with the ultimate goal of configurable and scalable execution of multiple DTs in heterogeneous computing environments.

The rest of this paper is organised as follows. An overview of the existing DTaaS platform is provided in Section~\ref{sec:dtaas}, including additional user needs and requirements, with the DTaaS extension presented in Section~\ref{sec:app}. A brief summary of related DT platforms is given in Section~\ref{sec:rel}. The final remarks are made in Section~\ref{sec:con}.

\section{DTaaS Platform Overview}\label{sec:dtaas}
\begin{figure}[htbp!]
    \centering
    \includegraphics[width=0.75\textwidth]{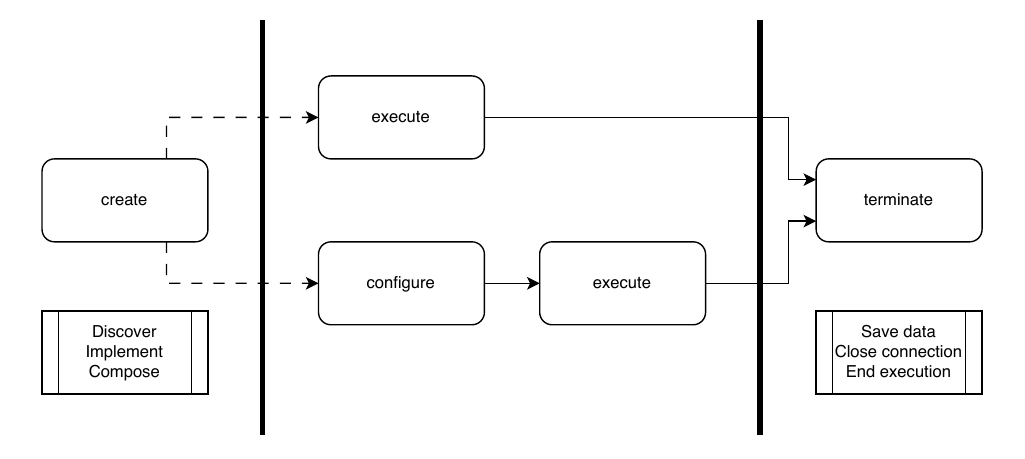}
    \caption{DT life-cycle phases. Dashed arrows indicate possible actions, whereas solid arrows indicate the taken action.}\label{fig:phases}
\end{figure}

The DT as a service (DTaaS) platform facilitates the development and management of DTs~\cite{talasila2024composable}, by providing users with means for (i) developing and sharing reusable DT components named DT assets, (ii) a development workspace to work on these assets, (iii) integrated platform services for communication, data storage and visualisation, and (iv) a structure of DT life-cycle phases.
In the context of DTaaS, assets can consist of models, data, functions, services, model solvers, each encapuslating parts of a DT that could be shared across other DT projects.
Users have access to private workspaces in which they can develop these DT assets, and trial run DTs there.
It is also possible to use private workspaces and create assets that are protected by Intellectual Property rights.
DTaaS enables users to define a set of life-cycle phases for any DT.
Figure~\ref{fig:phases} shown an example of DT life-cycle on the DTaaS platform. The \textit{create} phase: include all operations needed for the discovering the available DT assets on the platform, creation of a DT instance, e.g., selection of suitable OS execution environment, installing all required dependencies;
\textit{execute}: run a DT instance, establish bidirectional communication with PT, run DT simulators, integrate DT with visualisation tools, and so on;
\textit{reconfigure}: change DT configurations before executing the same, e.g., reconfigure monitors, change models used inside the DT, and so on. This step is only required if a user intends to receive different insights from the DT;
\textit{terminate}: save the state of the DT and support graceful shutdown of the DT, including termination operations that may affect the PT.
The user has control over which phases to include and their specification.

In general, a user can work with the platform in one of two ways.
The first consists in the user uploading to the platform a pre-existing and pre-packaged DT. 
In this case, the DT would consist of possibly tightly coupled services and models, ensuring the operation of the DT itself and the communication with the PT. 
Nevertheless, such coupling of models, and components in general would prevent reusability across projects -- even for the same user.
The second, and preferred way to work with the platform, which fully exploits its features, is to realize the DT by bringing together relevant available assets, such as data, models, functions, and tools.
These would have been made available for sharing previously,  by the same, or other users.

\subsection{User Needs \& Requirements}\label{sec:req}

DTaaS provides a structured way for dealing with the different lifecycle phases of DTs, as well as managing different kinds of assets associated with DTs, such as models, data, tools, etc.
Nevertheless, a few issues remain to be addressed, such as (i) how to build the models underlying DTs, and thereafter how to build DTs themselves, and (ii) how to support building DTs at the low level, i.e., enabling discovery, implementation, and reconfiguration of existing models.


In order to support resusability of assets across DT projects, it is crucial to provide storage and discovery mechanisms that enable users to locate those assets that are of interest to them.
The DTaaS platform provides a storage mechanism that allows users to store their assets locally, or on a remote server, such as a git server.
The latter is important, as it allows users to share their assets with other users across multiple instances of DTaaS.
In this setting, users have access to a git server from both within and outside of the DTaaS platform. The regular git workflows can be used to support (i) synchronisation of the DT assets to the user workspaces, (ii) develop assets outside of DTaaS and synchronize them to the platform assets, (iii) create new DTs from the user interface.

It is possible to differentiate DT assets into \textit{private} and \textit{common} categories. The private assets are visible only to owner while the common assets are visible to other users there by enabling asset-level collaboration.
Assuming that users finds relevant assets to incorporate in the new DT project, these are selected and used to compose the final DT.
Note that DT composition refers to (i) the coupling together of new assets developed from scratch, (ii) assets taken as-is from the asset repository (found via asset discovery), (iii) (re)configuring the assets taken from the asset library, (iv) creating configuration for DT.
At the end of this process, the user has succesfully created the DT, and the latter can be executed directly on the platform, or be subject to further reconfiguration. The platform allows the definition of termination mechanisms, with which it is possible to end the execution of the DT in a clean manner, while possibly saving the state and additional data as per user needs.
This process is summarized in Figure~\ref{fig:phases}.

\section{Federated Digital Twins}\label{sec:app}
\begin{figure}[tbh!]
    \centering
    \includegraphics[width=0.5\textwidth]{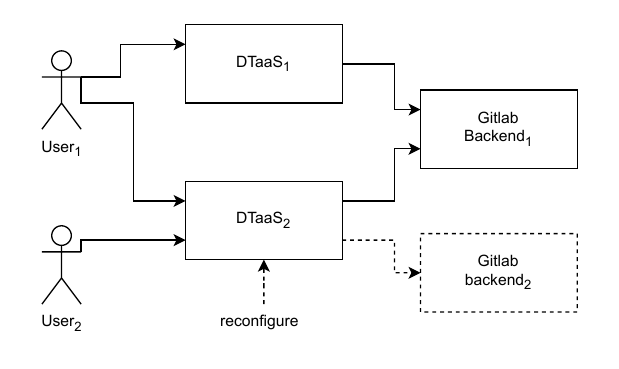}
    \caption{Collaboration across multiple DTaaS installations. The link from a DTaaS instance to Gitlab backend is configurable thus extending the collaboration across multiple DTaaS instances.} \label{fig:multiple-dtaas-instances}
\end{figure}

The DTaaS platform has been extended to support the collaboration between multiple users within and across multiple DTaaS platform instances as shown in Figure~\ref{fig:multiple-dtaas-instances}. This functionality represent the core contribution of this demo paper (the reader is pointed to the Resources section at the end of the article for links to relevant DTaaS resources).

As depicted in the figure, a superuser installs one instance of the DTaaS platform for many users, either locally or remotely on a web server.
Through each DTaaS instance, the corresponding user can discover existing assets -- those that have been made available for public use -- by querying the GitLab API (e.g., for the \texttt{GitLab Backend}$_1$ depicted in Figure~\ref{fig:multiple-dtaas-instances}).
This coupling allows users to make and push possible changes to the common assets, each through their own DTaaS platform instance. 
In addition, instances can be coupled to different Gitlab backends, depending on the concrete needs of the DTaaS users. 
Furthermore, users can have access to multiple instances of the platform, thus users are able to collaborate with different user groups.
\begin{figure}[htbp!]
    \centering
    \includegraphics[width=0.7\textwidth]{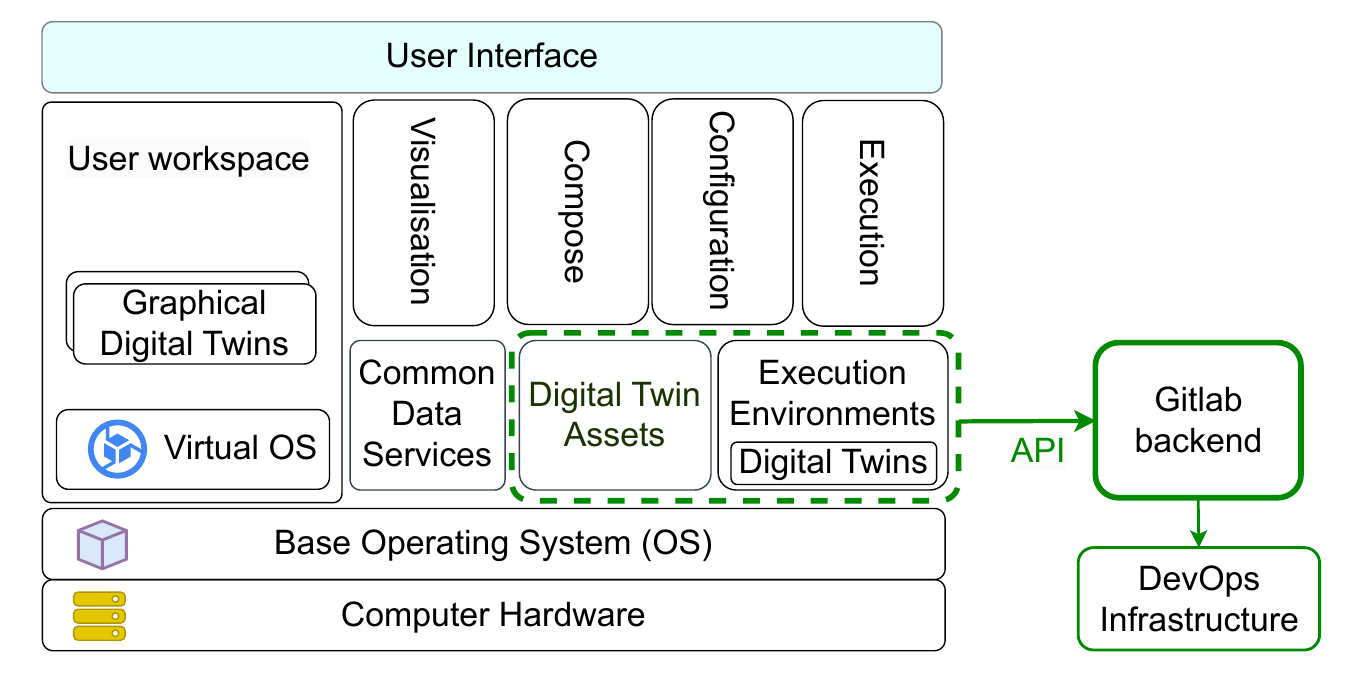}
    \caption{The layered architecture of the Digital Twin as a Service (DTaaS) and its dependence on Gitlab. The interaction uses Gitlab API.}
    \label{fig:dtaas-architecture}
\end{figure}
\begin{figure}[htbp!]
    \centering
    \includegraphics[width=0.7\textwidth]{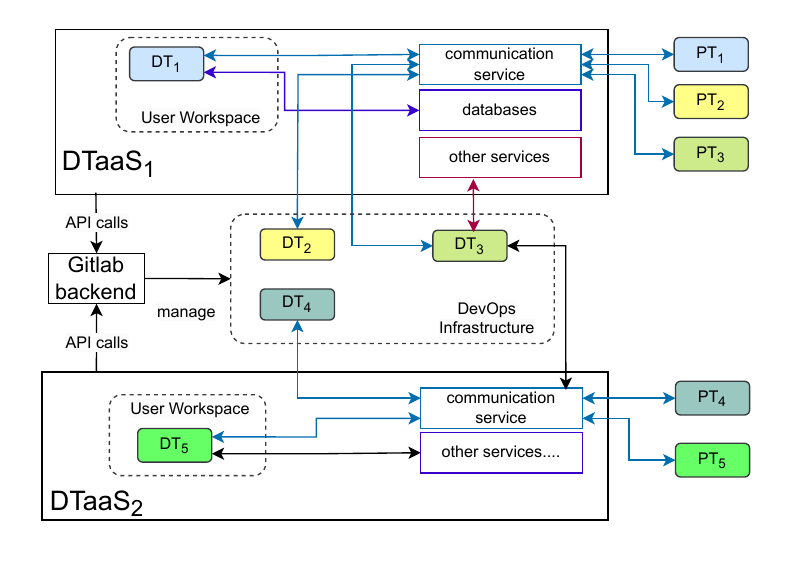}
    \caption{Snapshots of two DTaaS instances using a common Gitlab DevOps infrastructure.}\label{fig:snapshots}
\end{figure}
The layered DTaaS architecture and its dependence on Gitlab backends is depicted in Figure~\ref{fig:dtaas-architecture}. 
To compose a DT, reusble DT assets, as well as common data services (e.g., InfluxDB, MongoDB, RabbitMQ) should be present.
Different execution environments are supported, including docker, virtual machine, orKubernetes. 
The services such as DT Compose, Configuration, Execution and Visualisation are provided via the UI of the platform.
Some of these services (Compose, Configuration, Execution) are for managing the DTs while others (Visualisation) are services built on DTs for the users to gain insight during the operation of the PT.
The UI, provides a uniform abstraction to the user with or without the presence of a Gitlab backend.

Note that, different DTs could be contained within one instance, each having access to the different services offered by the platform.
The latter include communication services (e.g., RabbitMQ), enabling the exchange of data with the respective PTs, as well as databases such as InfluxDB, enabling users to build dashboards and gain insight over the performance of the PTs.

\subsection{DevOps on Digital Twin Platforms}
\label{subsec:devops}

The DevOps have proven to be useful for automated verification, validation and traceability in the Cyber-Physical Systems and DTs. The DTaaS uses the Gitlab backend to integrate DevOps onto the platform. The users of DTaaS are not exposed to the workings of Gitlab; instead, they are provided with a user interface that allows them to trigger the instantiation of DTs.

DTaaS uses the Gitlab DevOps infrastructure to support execution of DTs in two modes, namely continuous, and one off.
The most desirable execution patterns for a DT is the continuous execution. In this case, a DT instance is always running and is communicating with PT instance in either real-time or quasi real-time. 
The DevOps pipelines (of Gitlab) have traditionally been designed to support job execution that is triggered by events such as API calls, code commit, manual triggering etc.
The DTaaS uses the DevOps API of the Gitlab backend to instantiate the DT on DevOps infrastructure.
Well-established infrastructure configuration tools (including Ansible, Cloud formation, Terraform) can be used within Gitlab DevOps configuration files to achieve continuous execution of DTs.

The one off execution of DTs terminates within a time limit. The manager-worker frameworks support this kind of execution. 
The DTaaS provides a web interface on using which users can trigger one-off executions of DTs effortlessly on DevOps infrastructure. The execution results and logs are presented to the users.

Figure~\ref{fig:snapshots} shows the execution of DTs on the common Gitlab DevOps infrastructure. Two distinct installations of the DTaaS are linked to one Gitlab backend which in turn manages the DevOps infrastructure. The DTs not requiring desktop graphical interfaces are managed using the Gitlab backend. Each DT is either configured or reconfigured to make the DT run on a suitable infrastructure (such as docker, Kubernetes, etc.), to connect with common services and to connect with PT. This configuration can also be used to help DTs connect with external DTs, potentially belonging to another DTaaS instance. The $DT_3$ illustrates this scenario where one DT is managed by $DTaaS_1$ while it also interacts with common services and thereby the DT instances on $DTaaS_2$.
All PTs use at least one communication service for bidirectional interaction with their respective DTs.
In the example configuration shown in Figure~\ref{fig:snapshots}, $DTaaS_1$ owns $DT_1$ and $DT_2$, whereas 
$DTaaS_2$ owns $DT_4$ and $DT_5$. 
$DT_3$ is a federated DT that is managed by $DTaaS_1$ and communicates with $DTaaS_2$.
Federated DTs can be organised in different ways, e.g., composite, fleets, and hierarchical DTs~\cite{frasheri&25}.


\section{Related Work}\label{sec:rel}
The growing interest towards DTs and their development in the research community can be evidenced not only by the growing body of work on DTs, but also on the platforms and frameworks that have been proposed in the past years. 
A few survey papers provide comparisons of these frameworks and platforms, considering aspects such as system architecture, interoperability, scalability, reproduction of previous results, and composition~\cite{gil2024survey, ferko2023standardisation, ferko2022architecting}.
In contrast to existing solutions, DTaaS provides the infrastructure that enables users to orchestrate complementary services needed to execute and maintain DTs, while providing clear cut asset definitions and configurations.
Existing platforms can be run in conjuction with DTaaS, however the latter provides additional features compared to others, in terms of support for private worksapces, as well as enables asset sharing, thus fostering resusability. 
DTaaS tackles challenges related to the provision of support for different life-cycle phases as well service orchestration~\cite{gil2024survey,stojanovic2021methodology} by succesfully implementing software engineering practices such as microservices, DevOps, GitOps, and providing extra functionalities in terms of asset verification~\cite{reiterer2020continuous} and adoption of GitOps workflows~\cite{beetz2021gitops}.

\section{Conlusions}\label{sec:con}
In this demo paper we present an extension of the DTaaS platform that supports federated development and management of DTs and their related assets. 
This feature allows multiple users operating on different instances of DTaaS to run discovery of assets, as well push custom changes as needed by their projects through git APIs.
In the future, we will focus on developing rich case-studies to showcase the merits of DTaaS for a federated approach to DT development and management.

\section*{Resources}\label{sec:ref}

The software of the DTaaS platform are available at \url{https://github.com/INTO-CPS-Association/DTaaS}. The feature demo videos including DevOps features are available at\newline \url{https://into-cps-association.github.io/DTaaS/development/user/examples/index.html}.

\nocite{*}
\bibliographystyle{eptcs}
\bibliography{generic}
\end{document}